\documentclass{article}
\usepackage{amsfonts}
\usepackage{spconf,amsmath,graphicx}
\usepackage{float}

\usepackage{algorithm}
\usepackage{algpseudocode} 
\usepackage{amsmath, amssymb}

\usepackage{enumitem}
\setlist{nosep, leftmargin=14pt}

\usepackage{mwe} 

\makeatletter
\@ifundefined{FloatBarrier}{}{}
\makeatother

\title{Structure-Aware Adaptive Kernel MPPCA Denoising for Diffusion MRI}
\name{Ananya Singhal, Dattesh Dayanand Shanbhag, Sudhanya Chatterjee}
\address{GE HealthCare, Bengaluru, India}
\begin{document}
%
\maketitle
\begin{abstract}
Diffusion-weighted MRI (DWI) at high b-values often suffers from low signal-to-noise ratio (SNR), making image quality poor. Marchenko-Pastur PCA (MPPCA) is a popular method to reduce noise, but it uses a fixed patch size across the whole image, which doesn't work well in regions with different structures. To address this, we propose an adaptive kernel MPPCA (ak-MPPCA) that selects the best patch size for each voxel based on its local neighborhood. This improves denoising performance by better handling structural variations.
\end{abstract}
\section{Introduction}
\label{sec:intro}

DWI is crucial for characterizing tissue microstructure, but high b-value acquisitions suffer from low signal-to-noise ratios (SNR).
MPPCA has emerged as a popular approach for DWI denoising by leveraging random matrix theory to distinguish signal from noise \cite{veraart2016denoising}. Popularly this has been implemented in patch-based manner.
Recently, investigations have been published which aim at improving aspects of MPPCA denoising. Henriques et al. explored threshold-based criteria for component selection \cite{henriques2023efficient}. Several works have evaluated tensor-based approaches for multi-dimensional data denoising using MPPCA \cite{olesen2023tensor,herthum2024tensor}. The emphasis has been on methods to improve the noise estimate.
However, these methods employ fixed kernel sizes across entire volumes, which does not account for structural heterogeneity of the in-vivo image. 
Adaptive patching in MPPCA, as implemented in DESIGNER \cite{ades2018evaluation, chen2024optimization} attempts to overcome this challenge by selecting voxels for denoising based on both spatial proximity and signal similarity.

In this work, we introduce adaptive kernel MPPCA (ak-MPPCA), a denoising approach that estimates the optimal kernel size for each voxel based on the structural complexity of surrounding tissue. Structural complexity is quantified using image gradient information, allowing the method to adapt locally to anatomical variation. This eliminates the need for manual kernel size tuning and improves denoising performance by preserving fine structural details.

\section{Methods}
\label{sec:format}
MPPCA uses principles from random matrix theory to separate signal from noise in DWI data \cite{veraart2016denoising}. It removes noise-dominated eigenvalue components by identifying those that fall within the bounds of the Marchenko-Pastur distribution, thereby preserving meaningful signal. However, conventional implementations use a fixed kernel size across the image, which fails to adapt to local tissue heterogeneity. In this work, we propose a voxel-wise adaptive kernel estimation strategy that adjusts the kernel size for denoising based on the structural complexity of each voxel’s neighborhood.
\subsection{Kernel Map Estimation}\label{subsec:kernelmapestimation}
DWI data at a given b-value is typically represented as a 4D volume, where the fourth dimension corresponds to diffusion directions. In this section, we describe our method for estimating the optimal kernel size for each voxel to improve MPPCA-based denoising.
\subsubsection{Data Selection for kernel size estimation}\label{subsubsec:DataSel}
To estimate voxel-wise kernel sizes for denoising, we use the trace image from the lowest available b-value image, provided that $b\geq100$. This image is chosen because it is sufficiently diffusion-weighted to reflect tissue microstructure, while still preserving enough signal to capture structural complexity. This balance makes it suitable for guiding kernel size estimation.
\subsubsection{Gradient-Based Structural Characterization}
To assess structural heterogeneity around each voxel, we compute the image gradient information $(G)$ as follows:
\begin{equation}\label{eq:gradientmap}
G = \sqrt{\left(S_x * \left(I * G_\sigma\right)\right)^2 + \left(S_y * \left(I * G_\sigma\right)\right)^2}
\end{equation}
where $S_x$ and $S_y$ are the Sobel operators, $I$ is the image selected as described in Section~\ref{subsubsec:DataSel}, $G_\sigma$ is a Gaussian kernel with $\sigma = 4$, and $*$ denotes convolution. Gaussian smoothing reduces noise sensitivity during edge detection. Voxels with high gradient values indicate complex tissue structures, while low gradient values suggest simpler or more homogeneous regions. Here, $G$ has the same dimensions as $I$, a 3D volume corresponding to a diffusion direction in the DWI data (see Section~\ref{subsubsec:DataSel}).
\subsubsection{Kernel Size Determination}
This section describes the method used to estimate voxel-wise kernel sizes for MPPCA denoising based on image gradient information. Since gradient magnitudes are floating-point values and kernel sizes must be odd integers, we propose an algorithm to map gradient values to discrete kernel sizes. The user provides a lower $\left(k_l\right)$ and upper $\left(k_u\right)$ bound for kernel size, both as odd integers. Let $\{n_k\}_{k=1}^{K}$ denote the set of $K$ plausible kernel sizes within the range $[k_l, k_u]$.
Gradient values within the brain mask are clustered into $n_k$ groups using the k-means clustering algorithm. The brain mask is generated using median Otsu thresholding as implemented in DIPY~\cite{garyfallidis2014dipy}. Voxels outside the brain mask are assigned a fixed kernel size $k_h = H$. In our work, we consider $H = k_u$.
The clustered gradient image contains values $\left\{g_{n_1}, \ldots, g_{n_K}, g_h\right\}$. We assume that for clusters $g_{n_i}$ and $g_{n_j}$, if $i < j$, then $\left\{G\in g_{n_i}\right\} > \left\{G\in g_{n_j}\right\}$, $\forall \left\{i,j\right\}$. For each voxel at location $(x, y, z)$, if $G(x, y, z) = g$ and $g$ belongs to the $i^{\text{th}}$ cluster, then the assigned kernel size is $k_i$.\\
In summary, gradient values are grouped into clusters corresponding to the number of available kernel sizes. Voxels with higher gradient magnitudes receive smaller kernels, while those with lower gradients are assigned larger kernels. This results in a voxel-wise kernel size map with odd integer values, matching the dimensions of the gradient image. For this work, we consider kernel sizes to be isotropic. The algorithm has been shown in Algorithm \ref{alg:grad2kernel}.
%

\subsection{ak-MPPCA: MPPCA with estimated kernel size}
The MPPCA implementation is updated to consider the estimated kernel size for patch based denoising across the voxels. The remaining aspects of MPPCA denoising, such as noise estimation and signal thresholding is maintained as discussed in Veerart et al.~\cite{veraart2016denoising}. In this work, we set $k_l$ and $k_u$ to 5 and 9 respectively.

\begin{algorithm}[t]
	\caption{Mapping Gradient Magnitudes to Kernel Sizes for MPPCA}
	\label{alg:grad2kernel}
	\begin{algorithmic}[1]
		\Require Gradient magnitude image $G$ (refer Eq.~\ref{eq:gradientmap}); brain mask $M$; user-specified odd bounds $k_l, k_u$ with $k_l \le k_u$; allowable odd integer kernel sizes $\{n_k\}_{k=1}^{K} \subseteq [k_l, k_u]$
		\Ensure Kernel size map $\mathcal{K}$ (same size as $G$), odd and isotropic
		\State $\mathcal{K}_{\text{vals}} \gets \{\, k \in \mathbb{N} \mid k_l \le k \le k_u,\ k \text{ odd} \,\}$ \Comment{Allowed odd kernel sizes}
		\State $K \gets |\mathcal{K}_{\text{vals}}|$
		\State $H \gets k_u$ \Comment{Outside-brain kernel size}
		\State $\mathcal{S} \gets \{\, G(\mathbf{r}) \mid \mathbf{r} \in \Omega,\ M(\mathbf{r}) = 1 \,\}$ \Comment{Gradients inside brain}
		\State $\{\mathcal{C}_i\}_{i=1}^{K} \gets \text{k-means}(\mathcal{S}, K)$ \Comment{Cluster gradients into $K$ groups}
		\State $\mu_i \gets \frac{1}{|\mathcal{C}_i|} \sum_{g \in \mathcal{C}_i} g \quad \text{for } i=1,\dots,K$ \Comment{Cluster means}
		\State Compute permutation $\pi$ such that $\mu_{\pi(1)} \ge \mu_{\pi(2)} \ge \cdots \ge \mu_{\pi(K)}$ \Comment{Higher gradients first}
		\State Sort $\mathcal{K}_{\text{vals}}$ increasingly: $k_{(1)} < k_{(2)} < \cdots < k_{(K)}$ \Comment{Smaller kernels mapped to higher gradients}
		\ForAll{$\mathbf{r} \in \Omega$}
		\If{$M(\mathbf{r}) = 1$}
		\State $g \gets G(\mathbf{r})$
		\State Find $i \in \{1,\dots,K\}$ such that $g \in \mathcal{C}_{\pi(i)}$
		\State $\mathcal{K}(\mathbf{r}) \gets k_{(i)}$ \Comment{Monotonic mapping: high $G$ $\rightarrow$ small $k$}
		\Else
		\State $\mathcal{K}(\mathbf{r}) \gets H$ \Comment{Outside brain: assign $k_u$}
		\EndIf
		\EndFor
		\State \Return $\mathcal{K}$
	\end{algorithmic}
\end{algorithm}

\subsection{Data and Experiment}
All evaluations in this study were conducted using data from the Human Connectome Project (HCP) (MGH-USC)~\cite{fan2016mgh}. The proposed denoising method was applied to DWI datasets with $b$-values of 10000~$s/\text{mm}^2$ and 5000~$s/\text{mm}^2$ across three subjects. Comparative performance was assessed against publicly available implementations of MPPCA. Specifically, NYU-MPPCA refers to the implementation by Veraart et al.~\cite{veraart2016denoising}, DIPY-MPPCA refers to the version by Garyfallidis et al.~\cite{garyfallidis2014dipy}. The adaptive patching variant from the DESIGNER toolbox~\cite{chen2024optimization} was also included for fair comparison.
Both NYU-MPPCA and DIPY-MPPCA implementations require a fixed kernel size for denoising. In this study, we applied isotropic kernel sizes of 5, 7, and 9 for these two methods. For the proposed ak-MPPCA, the kernel size estimation limits were set to $[5, 9]$.

\begin{figure*}[!htbp]
	\centering
	\includegraphics[width=1.2\columnwidth]{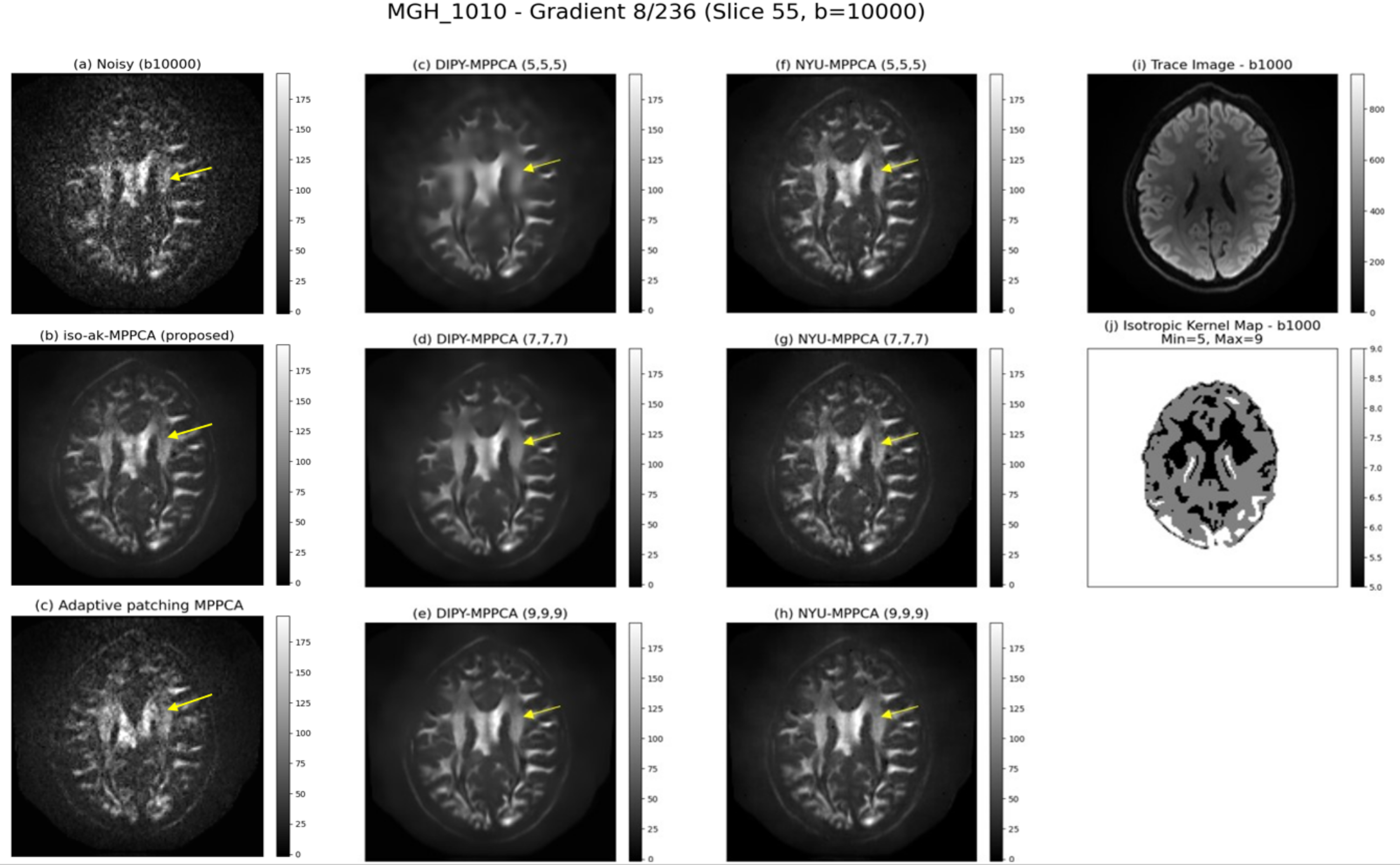}\\[1ex]
	\includegraphics[width=1.2\columnwidth]{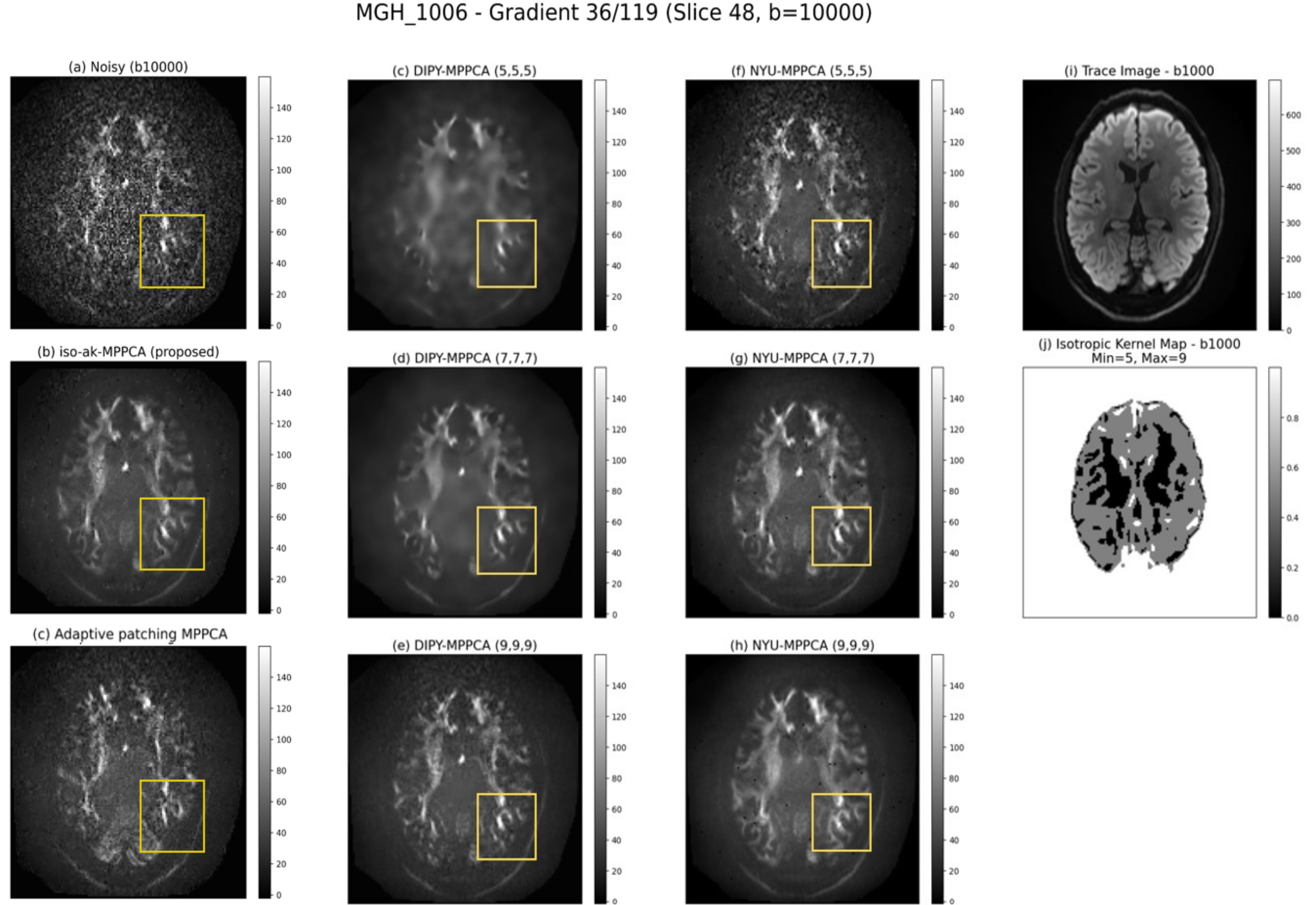}\\[1ex]
	\includegraphics[width=1.2\columnwidth]{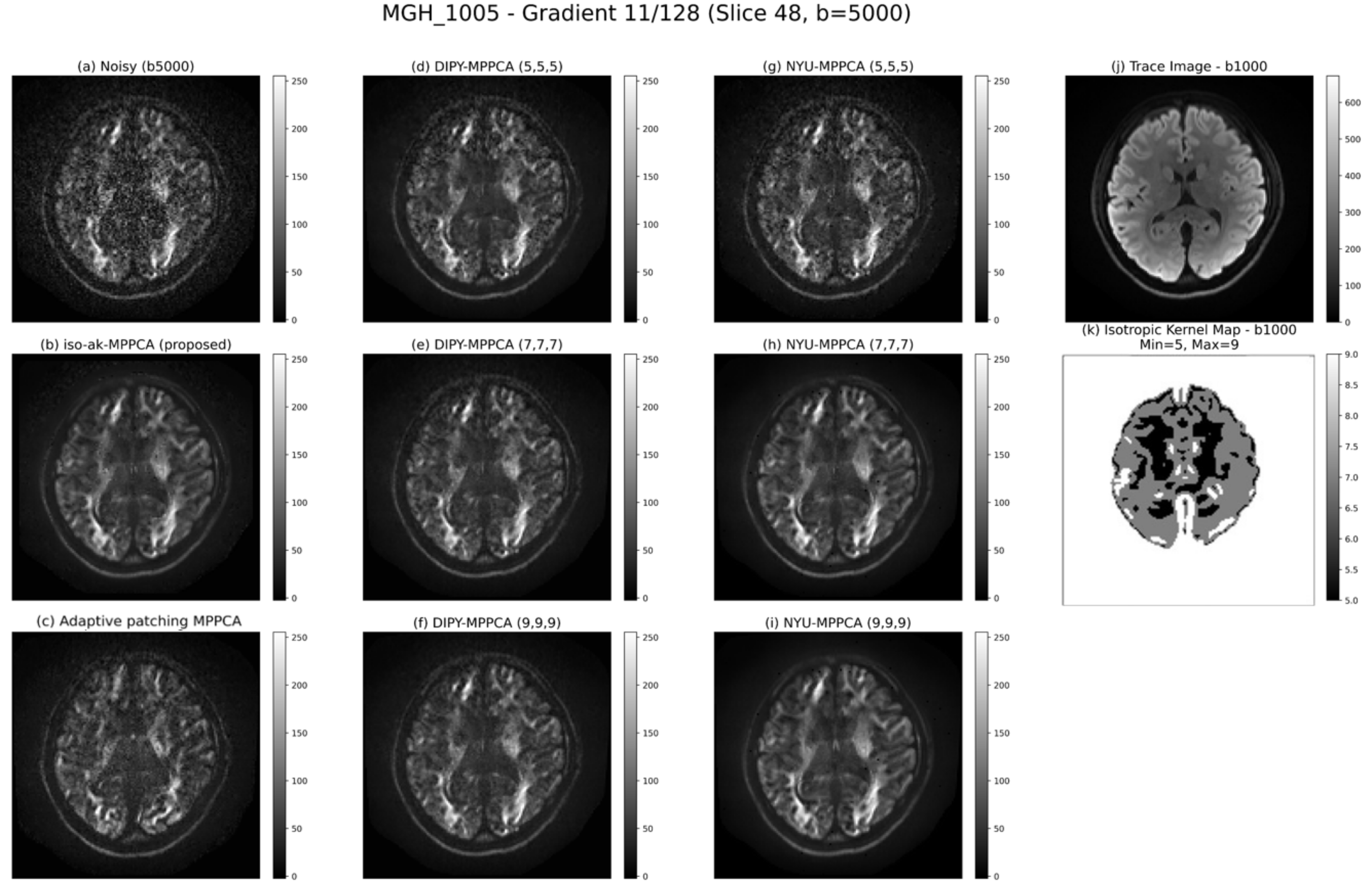}\\[1ex]
	\caption{Visual comparison of denoising results across methods for each subject. Panels show: (a) original noisy DWI image, (b) denoised image using the proposed ak-MPPCA method, (c) adaptive patching MPPCA~\cite{chen2024optimization}, (d–f) DIPY-MPPCA~\cite{garyfallidis2014dipy} with varying kernel sizes, (g–i) NYU-MPPCA~\cite{veraart2016denoising} with varying kernel sizes, (j) trace image used for kernel size estimation (see Section~\ref{subsubsec:DataSel}), and (k) voxel-wise kernel size map generated using Algorithm~\ref{alg:grad2kernel}. The kernel map in (k) was used to guide MPPCA denoising in (b). Yellow arrows/boxes highlight regions where the proposed method demonstrates improved preservation of structural detail and reduced blurring compared to other approaches (without requiring any kernel size adjustments).}
	\label{fig:combined}
\end{figure*}

\section{Results and Discussion}
\begin{figure}[!h]
	\begin{minipage}[b]{1.0\linewidth}
		\centering
		\centerline{\includegraphics[width=6.0cm]{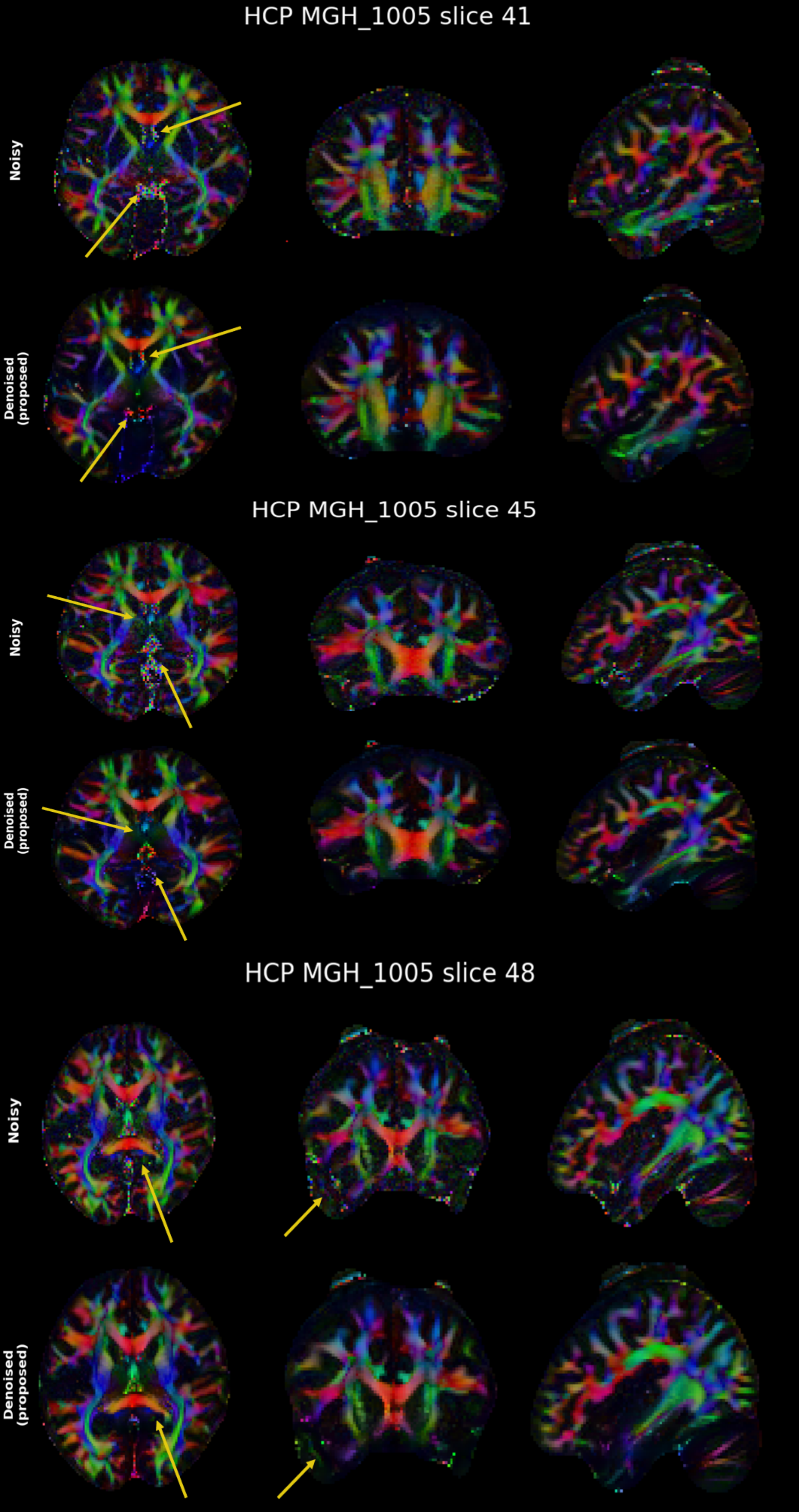}}
	\end{minipage}
	\caption{The primary diffusion directions for the noisy and ak-MPPCA denoised DWI data are shown color FA maps~\cite{pajevic1999color,garyfallidis2014dipy}. The proposed method improves the clarity of these maps.}
	\label{fig:res}
\end{figure}

Three subjects from the HCP (MGH-USC) dataset~\cite{fan2016mgh} were analyzed: MGH-1005, MGH-1006, MGH-1010. Denoised DWI results are shown in Figure~\ref{fig:combined}. For MGH-1010 and MGH-1006, DWI data with $b = 10000~s/\text{mm}^2$ was used, while for MGH-1005, results are shown for $b = 5000~s/\text{mm}^2$. The primary diffusion directions for the noisy and denoised (using proposed method) DWI data are represented using a red-green-blue representation to create a directionally encoded color (DEC) map (color FA maps) in Figure~\ref{fig:res} \cite{garyfallidis2014dipy, pajevic1999color}.
As shown in Figure~\ref{fig:combined}, the NYU-MPPCA and DIPY-MPPCA denoised DWI data depends on the kernel size choice (a hyperparameter to be optimized). In contrast, the proposed method estimates voxel-wise kernel sizes based on local structural information (refer Section~\ref{subsec:kernelmapestimation}). The corresponding kernel size maps are also shown in Figure~\ref{fig:combined}. These maps reveal that regions with high structural detail are assigned smaller kernel sizes, helping preserve fine features and reducing the risk of blurring during denoising.
Although the adaptive patching MPPCA~\cite{chen2024optimization} addresses structural blurring to some extent, it does not suppress noise as effectively. Among the compared methods, the proposed approach demonstrates superior performance in preserving anatomical structure while effectively reducing noise in DWI data.
As observed from the color FA maps in Figure~\ref{fig:res}, the proposed method improves the clarity of these maps, resolving noisy regions and preserving meaningful diffusion direction information.

\section{Conclusion}
We proposed ak-MPPCA, a denoising method for DWI that estimates voxel-wise kernel sizes based on local image gradients. Unlike fixed-size MPPCA methods~\cite{veraart2016denoising,garyfallidis2014dipy}, ak-MPPCA adapts to structural complexity, improving noise reduction and detail preservation. Evaluated on high b-value data from the HCP~\cite{fan2016mgh}, ak-MPPCA performed better than existing approaches, including adaptive patching~\cite{chen2024optimization}, in producing cleaner DWI and more accurate color FA maps.

\bibliographystyle{IEEEbib}
\bibliography{strings,refs}

\end{document}